\documentclass[final,3p,times,onecolumn]{elsarticle}
\usepackage{graphicx}
\usepackage{amsmath,amssymb}
\journal{Physics of the Dark Universe}
\begin{document}

\begin{frontmatter}

\title{Constraining theories of gravity by fundamental plane of elliptical galaxies}

\author[a,b,c,d]{Salvatore  Capozziello}
\ead{capozziello@unina.it}
\author[e]{Vesna  Borka Jovanovi\'{c}}
\author[e]{Du\v{s}ko Borka}
\author[f]{Predrag Jovanovi\'{c}}
\address[a]{Dipartimento di Fisica ''E. Pancini'', Universit\`{a} di Napoli ''Federico II'', Compl. Univ. di Monte S. Angelo, Edificio 6, Via Cinthia, I-80126, Napoli, Italy}
\address[b]{Istituto Nazionale di Fisica Nucleare (INFN) Sez. di Napoli, Compl. Univ. di Monte S. Angelo, Edificio 6, Via Cinthia, I-80126, Napoli, Italy}
\address[c]{Gran Sasso Science Institute, Viale F. Crispi, 7, I-67100, L'Aquila, Italy}
\address[d]{Laboratory for Theoretical Cosmology, Tomsk State University of Control Systems and Radioelectronics (TUSUR), 634050 Tomsk, Russia}
\address[e]{Department of Theoretical Physics and Condensed Matter Physics (020), Vin\v{c}a Institute of Nuclear Sciences, University of Belgrade, P.O. Box 522, 11001 Belgrade, Serbia}
\address[f]{Astronomical Observatory, Volgina 7, P.O. Box 74, 11060 Belgrade, Serbia}

\begin{abstract}
We show that fundamental plane of elliptical galaxies can be used  to obtain observational constraints on   metric theories of  gravity.  Being it  connected to global properties of ellipticals, it can fix   parameters of modified  gravity. Specifically, we  use fundamental plane to constrain modified theories of gravity with  Yukawa-like corrections which  commonly emerge in the  post-Newtonian limit. After giving  examples on how these corrections are derived,   we first analyze the velocity distribution of elliptical galaxies comparing theoretical results  of modified gravity with Yukawa-like corrections with astronomical data. According to these results, it is possible  to constrain the  parameters of the  corrections  discriminating  among classes of  models compatible with astronomical observations. We  conclude that fundamental plane can be used as a standard tool to   probe different theories of  gravity in the weak field limit.
\end{abstract}


\end{frontmatter}

\section{Introduction}
\label{sec1}

The need for  dark matter  emerged to describe 
dynamics of self-gravitating systems like
stellar clusters, galaxies, groups and clusters of galaxies since the 30's of last century\cite{Oort1,Zwicky}.  In all these cases, there is more matter
 than  that  accounted for by
luminous  components assuming the validity of Newton potential  at all astrophysical scales.

In order to explain such observational results, the first  considered possibility
 was  assuming the existence of  sub-luminous components  dubbed {\it dark matter}.  Many candidates
have been proposed  to supplement the missing matter but, up to now, there is no final indication for their existence both at fundamental and astrophysical level \cite{revdonne}. 

Furthermore, cosmological observations require  another
unknown dynamical component, the so-called  {\sl dark energy}, to account for the
current accelerated expansion of the universe
\cite{Sullivan:2011kv,Ade:2013kta,planck}.  Also in this case, addressing the phenomenon at particle level is revealing a severe challenge.  Hence,  the
need of imposing unknown dark  components 
could be nothing else but the  signal of the breakdown of General Relativity
 at galactic, extragalactic and cosmological scales.  In this context, modified  theories of
gravity, like $f(R)$, Brans-Dicke, and Gauss-Bonnet gravity,  could be a way  to explain
cosmic accelerated expansion, large scale structure and galactic dynamics   \cite{capo02,clif05,PRnostro,reviewodi,reviewodi1,reviewmauro,reviewvalerio,libro,oikonomou, Lobo,nojiri}.

In  the weak-field limit,  alternative  theories of
gravity are expected to reproduce  General Relativity 
which is  successfully tested  at Solar System scales \cite{Will93}.  
Several  proposals have been
formulated to  go beyond  $\Lambda$CDM in view of 
possible alternatives to explain astrophysical and
cosmological observations. In particular,  the flat rotation curves of
galaxies,  can be addressed by  MOdified Newtonian
Dynamics (MOND) \cite{Milgrom:1983ca}  which adopts  an acceleration scale to account for
 high velocities without  dark matter. Besides  phenomenology, this acceleration scale could be some new  fundamental parameter of nature \cite{bern11, bern2}. 
 
 MOND has been relativistically improved  by TeVeS \cite{Bekenstein:2004ne}, an approach  including  additional
vector and scalar gravitational degrees of freedom  to the standard tensor field of General Relativity.   MOND and TeVeS have been tested  by  gravitational lensing
 concluding  that  non-trivial  dark matter components has to be added to match exactly 
astrophysical  observations \cite{Ferreras:2007kw, Mavromatos:2009xh,Ferreras:2009rv,Ferreras:2012fg}. 

 MOND and TeVeS are two prototypes of   several extensions and modifications of General Relativity  proposed to work at  infrared scales. For example, $f(R)$ gravity is a natural extension of Einstein's theory: it does not fix a priori the form of the gravitational action, like the Einstein-Hilbert one, but assumes that it can be reconstructed by observations \cite{PRnostro}. In this picture, dark components are a sort of curvature fluid  acting as an  interaction field at astrophysical and cosmological scales \cite{capo12,mantica}.  
 
 Besides $f(R)$, more detailed theories have been proposed to address the phenomenology. However, some common features can be put in evidence for any proposal aimed to explain dynamics without dark component: $i)$ General Relativity has to be reproduced at certain scale (e.g. at Solar System); $ii)$ further degrees of freedom reduce to corrections of the Newtonian potential \cite{Schmidt}; $iii)$ these corrections are often Yukawa-like terms characterizing the scale where the behavior begins to stand out with respect to  the  Newtonian dynamics. 
 
 As reported in \cite{capo15}, several  metric theories shows Yukawa-like corrections.   Clearly, as we will show below,  their  parameters strictly depend on the specific gravitational   theory worked out in the weak field limit.   In some sense, fixing them means selecting the theory of gravity. However, the limitation of this paradigm is that the given theory of gravity have to be developed in the post-Newtonian regime and strong field effects are not considered. 
 
Finally, this kind of corrections has been often investigated at Solar System, local and microscopic scales (the so called "fifth-force" issue \cite{fifth} ). On the other hand, a systematic investigation at galactic and extragalactic scales has never been performed, as far as we know.

In this paper, we  propose a new approach by which the weak field limit of metric theories can be systematically investigated at extragalactic scales. To this aim, we shall  adopt the fundamental plane (FP) of galaxies. The final goal is fixing the values of  correction parameters with respect  to the standard Newtonian potential, in order to select the corresponding field theory. Here, we perform the analysis assuming Yukawa-like corrections  because  wide classes of theories show them in the weak-field limit. In any case, the protocol can be adopted also for other gravitational corrections.

 Let us start from the  empirical fact  that some global properties of normal elliptical galaxies are correlated. It is well known that there are three main global observables: the central projected velocity dispersion  $\sigma_0$, the effective radius $r_e$, and the mean effective surface brightness (within $r_e$) $I_e$ \cite{bork16,bork19}. Any of the three parameters may be estimated from the other two, and, varying them,    a plane is described  within a more general three-dimensional configuration space. This correlated plane is  referred to as the FP. It is defined and discussed in  detail in several papers, see e.g \cite{bend92,bend93,busa97,saul13,tara15} and references therein.

This important empirical relation \cite{busa97}:

\begin{equation}
log(r_e) = a \times log(\sigma_0) + b \times log(I_e) + c,
\label{equ11}
\end{equation}
gives us the possibility to obtain  observational constraints on the structure, formation, and evolution of early-type galaxies. Here we shall adopt Eq.\eqref{equ11} to constrain parameters of  gravity theories. It is worth noticing that a FP can be defined for several self-gravitating systems ranging from stellar clusters up  to gamma ray bursts \cite{Dainotti}.

The content of this paper is as follows. In Section 2,  we  discuss   Yukawa-like corrections  emerging from  modified theories of gravity. In particular, we report the standard case of $f(R)$ gravity where Yukawa-like corrections naturally emerge. In Table I, we show several examples of theories where Yukawa corrections are derived. In Section 3,  we recover the FP of elliptical galaxies by  theories of gravity with Yukawa-like corrections, describing the  observations and giving  details on our approach. Constraints on  Yukawa  parameters by FP are  considered in   Section 4. Section 5 is devoted to  draw  conclusions.

\section{Yukawa-like corrections in the weak field limit of metric theories}
\label{sec2}
As shown in several studies, Yukawa-like corrections emerges as common  features for several metric theories of gravity. In particular for theories containing higher-order curvature invariants or  scalar-tensor theories. See \cite {libro} for a comprehensive review. As an example, we will show here the well-known case of $f(R)$ gravity which can be considered  as a sort of paradigm in this sense.

Let us start from the Einstein-Hilbert   action improved with a generic analytic function of the Ricci scalar $R$, that is  $f(R)$ gravity. The field equations  are 
\begin{equation}\label{fe}
f'(R)R_{\mu\nu}-\frac{1}{2}f(R)g_{\mu\nu}-f'(R)_{;\mu\nu}+g_{\mu\nu}\Box
f'(R)=\kappa T_{\mu\nu}\,,
\end{equation}
with the  trace  
\begin{equation}\label{fetr}
3\Box f'(R)+f'(R)R-2f(R)= \kappa T \,.
\end{equation}
Here $\kappa$ is the gravitational coupling and $T_{\mu\nu}$ is the energy-momentum tensor of matter.
We are interested in external solutions so we can ignore the matter contribution.

In the weak field limit, we can perturb the metric tensor with respect to the Minkowski background, i.e. 
$g_{\mu\nu}\,=\,\eta_{\mu\nu}+h_{\mu\nu}$ with $|\eta_{\mu\nu}|\ll |h_{\mu\nu}|$. Let us   assume an analytic 
 $f(R)$ Lagrangian expandable in Taylor series 
\begin{eqnarray}\label{sertay}
f(R)=\sum_{n}\frac{f^n(R_0)}{n!}(R-R_0)^n\simeq
f_0+f'_0R+f''_0R^2+f'''_0R^3+...
\end{eqnarray}
where the prime indicates derivatives with respect to $R$.
The field equations,  in the post-Newtonian limit up to ${\mathcal O}(4)$ order, are
\begin{eqnarray}\label{eq2}
&&f'_0rR^{(2)}-2f'_0g^{(2)}_{tt,r}+8f''_0R^{(2)}_{,r}-f'_0rg^{(2)}_{tt,rr}+4f''_0rR^{(2)}=0\,,\nonumber\\
&&f'_0rR^{(2)}-2f'_0g^{(2)}_{rr,r}+8f''_0R^{(2)}_{,r}-f'_0rg^{(2)}_{tt,rr}=0\,,\nonumber\\
&&2f'_0g^{(2)}_{rr}-r\left[f'_0rR^{(2)}-f'_0g^{(2)}_{tt,r}-f'_0g^{(2)}_{rr,r}+4f''_0R^{(2)}_{,r}+4f''_0rR^{(2)}_{,rr}\right]=0\,,\nonumber\\
&&f'_0rR^{(2)}+6f''_0\left[2R^{(2)}_{,r}+rR^{(2)}_{,rr}\right]=0\,,\nonumber\\
&&2g^{(2)}_{rr}+r\left[2g^{(2)}_{tt,r}-rR^{(2)}+2g^{(2)}_{rr,r}+rg^{(2)}_{tt,rr}\right]=0\,.
\end{eqnarray}
These equations can be  integrated giving the  solution
\begin{align}
\label{mesol} g_{tt}(r)&=\,1-\frac{GM}{f'_0r}+\frac{\delta_1 L^2 e^{-r/L}}{3}\,,\\
\label{mesol1} g_{rr}(r) &= 1+\frac{GM}{f'_0r}+\frac{\delta_1L^2(1+r/L)e^{-r/L }}{3r}\,,\\
R\,&=\,\frac{\delta_1e^{-r/L}}{r}\,.
\end{align}
where $g_{tt}$ and $g_{rr}$ are  the time and radial  metric coefficient respectively. $R$ is  Ricci scalar derived for this problem.  We are assuming  the spherical symmetry. Here $L\,\doteq\,\sqrt{-6f''_0/f'_0}$ is a scale length. The constants $\delta_1$ and $f'_0$ can be combined into another constant $\delta$ giving the strength of the correction  \cite{capo12,yukawa1,yukawa2,yukawa3,card11,Napolitano2012,Cap-def-Sal2009}. 
Being 

The functions  $g_{tt}$ and $g_{rr}$ give two gravitational potentials 
\begin{eqnarray}
\label{eq:PHI}\Phi(r) &=& -\frac{2G M \left(1+\delta   e^{-\frac{r}{L}}\right)}{rc^2(\delta +1)}, \\
\label{eq:PSI}\Psi(r) &=& \frac{2G M }{rc^2}\left[\frac{\left(1+\delta  e^{-\frac{r}{L}}\right)}{(\delta +1)}+\frac{\left(\frac{\delta  r
   e^{-\frac{r}{L}}}{L}-2\right)}{(\delta +1)}\right]\,, 
\end{eqnarray}
being, in general, 
\begin{equation}
\label{Edd}
    g_{tt}=1-\frac{\Phi(r)}{c^2}\,,\;\;\;\;\;\;  g_{rr}=1+\frac{\Psi(r)}{c^2}\,.
     \end{equation}
Eqs.\eqref{eq:PHI} and\eqref{eq:PSI} coincide  as soon as $r\rightarrow\infty$, i.e. asymptotically they give $\Phi\simeq\Psi$ that is the Newtonian potential is recovered.  With a suitable readjustment of parameters, 
\begin{equation}
G\longrightarrow \frac{2G}{1+\delta}\,,\; \;\;\;\;\alpha \longrightarrow \frac{\delta}{\delta+1}\,,\; \;\;\;\;L\longrightarrow\frac{1}{\lambda}\,,
\end{equation}
and for $c=1$, Eqs.(\ref{eq:PHI}) and (\ref{eq:PSI}) can be adapted to  the standard form 
\begin{equation}
\boxed{\Phi(r) = -\dfrac{GM(r)}{r}\left[1 + \alpha \exp(-\lambda r) \right]}\;\;.
\label{equ21}
\end{equation}
Again $\lambda=L^{-1}$ is a scale length and $\alpha$ gives the strength of the correction
\cite{capo15,card11,bork13}. It is worth noticing that $\displaystyle{L=\lambda^{-1}=\frac{h}{m_R\, c}}$ is a Compton length which can be related to an effective mass $m_R$ coming from  curvature. In early universe cosmology, it is the so-called    {\it Starobinsky scalaron} \cite{star80}.

A potential like  Eq. (\ref{equ21}) is very generic for a large class of theories. Various combinations of Yukawa-like terms intervene in several metric theories of gravity.  In Table 1, we report some examples where higher-order curvature terms or scalar fields give this kind of corrections in the weak field limit.  The paradigm is that, adding further degrees of freedom in the Einstein-Hilbert action, the typical outcome is one or more Yukawa-like corrections in the post-Newtonian limit. In some sense, the exception is General Relativity where only the standard Newtonian gravitational potential is recovered.

With this considerations in mind, we will study the FP and constrain  $\lambda$ and $\alpha$ by astronomical data. These parameters can suitably reproduce dark matter effects \cite{capo12}. According to this procedure, one can, in principle,  ''reconstruct''  the class of theories of gravity ''compatible'' with observations. In some sense, this is an "inverse scattering" procedure which could reveal  useful to probe theories by galactic data. It is a sort of   ''blind'' approach in which we are not requesting a priori the validity of a given model but we are asking for classes of compatible theories fixed by the range of parameters.

\begin{table}
\begin{tabular}{c|c|cl}
\hline
  Modified  Gravity Model  & Corrected Newtonian potential & Yukawa parameters \\
\hline &&&
\\
 $f(R)$ & $\Phi(r)=-\frac{GM}{r}\biggl[1+\alpha\,e^{-m_R r}\biggl]$ & $\begin{array}{ll}{m_R}^2\,=\,-\frac{f_R(0)}{6f_{RR}(0)}\end{array}$ \\
\hline
 & & &\\
$f(R,\Box R)= R+ a_0R^2+a_1R\Box R$ & $\Phi(r)=-\frac{GM}{r}\left(1+c_0e^{(-r/l_0)}+c_1e^{(-r/l_1)}\right)$ & $\begin{array}{ll}{c_{0,1}}\,=\,\frac{1}{6}\mp\frac{a_0}{2\sqrt{9a_0^2+6a_1}}\\\\  l_{0,1}\,=\,\sqrt{-3a_0\pm\sqrt{9a_0^2+6a_1}}\end{array}$ \\
\hline &&&
\\
$f(R,\Box R,..\Box^k R)= R+\Sigma_{k=0}^p a_kR\Box^k R$ & $\Phi(r)=-\frac{GM}{r}\left(1+\Sigma_{k=0}^{p}c_i\exp(-r/l_i)\right)$ & $c_i\,,\;l_i$ are functions of $a_k$. See \cite{Schmidt}. \\
\hline
 & & &\\
   $f(R,\,R_{\alpha\beta}R^{\alpha\beta})$ & $\Phi(r)=-\frac{GM}{r}\biggl[1+\frac{1}{3}\,e^{-m_R r}-\frac{4}{3}\,e^{-m_Y r}\biggl]$ & $\begin{array}{ll}{m_R}^2\,=\,-\frac{1}{3f_{RR}(0)+2f_Y(0)} \\\\{m_Y}^2\,=\,\frac{1}{f_Y(0)}\,,\, Y=R_{\mu\nu}R^{\mu\nu}
\end{array}$ \\
\hline  
& & & \\  
$f(R,\,\cal{G})$\,, \; ${\cal G}= R^2-4R_{\mu\nu}R^{\mu\nu}+ R_{\alpha\beta\mu\nu}R^{\alpha\beta\mu\nu}$& $\Phi(r)=-\frac{GM}{r}\biggl[1+\frac{1}{3}\,e^{-m_1 r}-\frac{4}{3}\,e^{-m_2 r}\biggl]$ & $\begin{array}{ll}{m_1}^2\,=\,-\frac{1}{3f_{RR}(0)+2f_Y(0)+2f_Z(0)} 
\\\\{m_2}^2\,=\,\frac{1}{f_Y(0)+4f_Z(0)}\,,\,Z=R_{\alpha\beta\mu\nu}R^{\alpha\beta\mu\nu}
\end{array}$\\
\hline  & & & 
\\  
$f(R,\,\phi)+\omega(\phi)\phi_{;\alpha}\phi^{;\alpha}$ &
$\begin{array}{ll}\Phi(r)=-\frac{GM}{r}\biggl[1+g(\xi,\eta)\,e^{-m_R\tilde{k}_R r}+\\\\\qquad\qquad+[1/3-g(\xi,\eta)]\,e^{-m_R\tilde{k}_\phi r}\biggr]\end{array}$
&
$\begin{array}{ll}{m_R}^2\,=\,-\frac{1}{3f_{RR}(0,\phi^{(0)})}\\\\{m_\phi}^2\,=\,-\frac{f_{\phi\phi}(0,\phi^{(0)})}{2\omega(\phi^{(0)})}\\\\\xi\,=\,\frac{3{f_{R\phi}(0,\phi^{(0)})}^2}{2\omega(\phi^{(0)})}\\\\\eta\,=\,\frac{m_\phi}{m_R}\\\\g(\xi,\,\eta)\,=\,\frac{1-\eta^2+\xi+\sqrt{\eta^4+(\xi-1)^2-2\eta^2(\xi+1)}}{6\sqrt{\eta^4+(\xi-1)^2-2\eta^2(\xi+1)}}\\\\{\tilde{k}_{R,\phi}}^2\,=\,\frac{1-\xi+\eta^2\pm\sqrt{(1-\xi+\eta^2)^2-4\eta^2}}{2}
\end{array}$ \\
\hline & & & 
\\  
$f(R,\,R_{\alpha\beta}R^{\alpha\beta},\phi)+\omega(\phi)\phi_{;\alpha}\phi^{;\alpha}$
&
$\begin{array}{ll}\Phi(r)=-\frac{GM}{r}\biggl[1+g(\xi,\eta)\,e^{-m_R\tilde{k}_R\,r}+\\\\\,\,\,\,+[1/3-g(\xi,\eta)]\,e^{-m_R\tilde{k}_\phi\,r}-\frac{4}{3}\,e^{-m_Y r}\biggr]\end{array}$
&
$\begin{array}{ll}{m_R}^2\,=\,-\frac{1}{3f_{RR}(0,0,\phi^{(0)})+2f_Y(0,0,\phi^{(0)})}\\\\{m_Y}^2\,=\,\frac{1}{f_Y(0,0,\phi^{(0)})}\\\\{m_\phi}^2\,=\,-\frac{f_{\phi\phi}(0,0,\phi^{(0)})}{2\omega(\phi^{(0)})}\\\\\xi\,=\,\frac{3{f_{R\phi}(0,0,\phi^{(0)})}^2}{2\omega(\phi^{(0)})}\\\\\eta\,=\,\frac{m_\phi}{m_R}\\\\g(\xi,\,\eta)\,=\,\frac{1-\eta^2+\xi+\sqrt{\eta^4+(\xi-1)^2-2\eta^2(\xi+1)}}{6\sqrt{\eta^4+(\xi-1)^2-2\eta^2(\xi+1)}}\\\\{\tilde{k}_{R,\phi}}^2\,=\,\frac{1-\xi+\eta^2\pm\sqrt{(1-\xi+\eta^2)^2-4\eta^2}}{2}
\end{array}$ \\
\hline 

\end{tabular}
\caption{Yukawa-like corrections are a general feature of several modified gravity models. In particular,  they emerge in Extended Theories of Gravity which are natural extension of General Relativity \cite{PRnostro}. In some sense, further degrees of freedom, related to  higher-order terms or scalar fields, give rise to these corrections in the weak field limit. This is a general result as discussed in  \cite{Schmidt}. In the Table, we report examples of modified gravity models showing  Yukawa-like corrections in the post-Newtonian limit. Detailed discussions of these results   are reported in \cite{nojiri, Schmidt, capo15, revelles, stabile}.}
\end{table}

\section{Fundamental plane of modified gravity with  Yukawa-like corrections}
\label{sec3}
As said above, we are going to constrain the  parameters $\alpha$ and $\lambda$ considering a sample of ellipticals  that distribute  along the FP. We shall describe the observations and the method that we are going to adopt.

\subsection{Observations and method}

We use the observational data for physical properties of stellar systems given in the paper by Burstein et al. (1997) \cite{burs97}. The table ''Global Relationships for Physical Properties of Stellar Systems'', given in ASCI format and labeled 'metaplanetab1', is available within the arXiv version of \cite{burs97}: https://arxiv.org/e-print/astro-ph/9707037. It summarizes data, like the self-consistent effective radii, effective luminosities, characteristic dynamic velocities, and some other related data. In our previous paper \cite{bork16}, we  already described the columns from this table which are  interesting in the present case.  We  also  emphasize, that from the sample of 1150 observed galaxies, we selected and studied 401 ellipticals.

Let us  take into account  the relation for circular velocity, consisting of the Newtonian contribution, and  the correction term due to modified gravity. Then, for the given Yukawa  parameters ($\alpha$, $\lambda$), we calculate the theoretical values of velocity dispersion (see  the next Subsection 3.2).

\subsection{Velocity dispersion and the singular isothermal sphere model}

In the case of Newtonian potential,  it is: $\Phi_N(r) = -\dfrac{GM(r)}{r}$ and circular velocity $v_N^{2}(r) = r \cdot \Phi_N^\prime(r)$ \cite{capo15}. In the case of modified potential, supposing the spherically distributed mass in elliptical galaxies, we have $v_c^{2}(r) = r \cdot \Phi^\prime(r)$ \cite{capo07}. Here we start from the Yukawa-like gravitational potential \eqref{equ21} and derive the connection between $v_c^{2}(r)$ and parameters of this potential. Let us start from Eq. (\ref{equ21}) and its derivative. We have:
\begin{eqnarray}
\Phi(r) &=& -\dfrac{GM(r)}{r} - \alpha \dfrac{GM(r)}{r} e^{-\lambda r}; \nonumber \\[0.2cm]
\Phi^\prime(r) &=& \left(-\dfrac{GM(r)}{r}\right)^\prime + \alpha \left(-\dfrac{GM(r)}{r}\right)^\prime e^{-\lambda r} 
                +\alpha \left(-\dfrac{GM(r)}{r}\right) \left(e^{-\lambda r}\right)^\prime \nonumber \\[0.2cm]
 &=& \Phi_N^\prime(r) + \alpha e^{-\lambda r} \Phi_N^\prime(r) + \alpha \Phi_N(r) (-\lambda) e^{-\lambda r} \nonumber \\[0.2cm]
r\Phi^\prime(r) &=& r\Phi_N^\prime(r) + \alpha e^{-\lambda r} r\Phi_N^\prime(r) - \alpha \lambda e^{-\lambda r} r\Phi_N(r) \nonumber \\[0.2cm]
v_c^{2}(r) &=& v_N^{2}(r) + \alpha e^{-\lambda r} v_N^{2}(r) + \alpha \lambda re^{-\lambda r} v_N^{2}(r) \nonumber \\[0.2cm]
 &=& \dfrac{GM(r)}{r} + \dfrac{GM(r)}{r} \alpha \left(1 + \lambda r\right)e^{-\lambda r}.
\label{equ32}
\end{eqnarray}

\noindent Therefore the squared circular velocity $v_c^{2}(r) = r \cdot \Phi^\prime(r)$ takes the form: $v_c^{2}(r) = \dfrac{GM(r)}{r}\left(1 + \alpha \left(1 + \lambda r \right) e^{-\lambda r} \right)$. Now, we can write this expression as a sum of the Newtonian contribution $v_N^{2}(r)$ and the correction term due to modified gravity $v_{corr}^{2}(r)$:

\begin{equation}
v_c^{2}(r) = v_N^{2}(r) + v_{corr}^{2}(r),
\label{equ33}
\end{equation}

\noindent where:
\begin{eqnarray}
v_N^{2}(r) &=& \dfrac{GM(r)}{r} \nonumber \\[0.2cm]
v_{corr}^{2}(r) &=& \dfrac{\alpha GM(r) \cdot \left(1 + \lambda r \right)}{r} e^{-\lambda r}.
\label{equ34}
\end{eqnarray}

\noindent Here, it is important to stress that the corrective velocity term can be expressed also via Newtonian velocity dispersion as $v_{corr}^{2}(r) = \alpha v_N^{2}(r) (1 + \lambda r) e^{-\lambda r}$.

For the mass distribution in elliptical galaxies, we can assume the singular isothermal sphere (SIS) model. Then, the density profile is $\rho_{SIS}(r) = \dfrac{\sigma_{SIS}^2}{2\pi Gr^2}$, and the corresponding mass, included within a radius $r$, grows linearly with $r$ as:

\begin{equation}
M_{SIS}(r) = \dfrac{2\sigma_{SIS}^2}{G}\cdot r.
\label{equ35}
\end{equation}

\noindent Using Eq. (\ref{equ35}), for the circular velocity, we get:

\begin{equation}
v_c^2(r) = 2\sigma_{SIS}^2 \left( 1 + \alpha \left( 1 + \lambda r \right) e^{-\lambda r} \right),
\label{equ36}
\end{equation}

\noindent and then, taking into account Eq. (\ref{equ34}), we have:

\begin{eqnarray}
v_N^2(r) &=& 2\sigma_{SIS}^2 \nonumber \\[0.2cm]
v_{corr}^2(r) &=& 2\alpha \sigma_{SIS}^2 \left( 1 + \lambda r \right) e^{-\lambda r}.
\label{equ37}
\end{eqnarray}
For the considered sample of elliptical galaxies, the Newtonian circular velocity at the effective radius is \citep[see the explanation of Table 1 in Ref. ][]{burs97} $v_N(r_e) = \sigma_0$, where $\sigma_0$ is the observed velocity dispersion. Therefore,

\begin{equation}
\sqrt{2}\sigma_{SIS} = \sigma_0.
\label{equ38}
\end{equation}
From Eqs. (\ref{equ36}) and (\ref{equ38}),  we have:

\begin{equation}
v_c^2(r) = \sigma_0^2 \left( 1 + \alpha \left( 1 + \lambda r \right) e^{-\lambda r} \right),
\label{equ39}
\end{equation}

\noindent and,  furthermore, 

\begin{eqnarray}
v_N^2 &=& \sigma_0^2 \nonumber \\[0.2cm]
v_{corr}^2 &=& \alpha \sigma_0^2 \left( 1 + \lambda r \right) e^{-\lambda r}.
\label{equ310}
\end{eqnarray}
The circular velocity at the effective radius, i.e. for $r = r_e$, is:

\begin{equation}
v_c^2(r_e) = \sigma_0^2 \left( 1 + \alpha \left( 1 + \lambda r_e \right) e^{-\lambda r_e} \right).
\label{equ311}
\end{equation}

\noindent For the sake of simplicity, we introduce the new variable:

\begin{equation}
w = \lambda r_e,
\label{equ312}
\end{equation}

\noindent and, from Eq.(\ref{equ311}),  it is:

\begin{equation}
v_c^2(r_e) = \sigma_0^2 \left( 1 + \alpha \left( 1 + w \right) e^{-w} \right).
\label{equ313}
\end{equation}
As in the Newtonian case where $v_N(r_e) = \sigma_0$ (see Eq. (\ref{equ310})), we can assume that the expression $v_c(r_e) = \sigma^{theor}$ is valid also in the case of modified gravity with Yukawa-like corrections, where $\sigma^{theor}$ is a theoretical velocity dispersion. Therefore:

\begin{equation}
\boxed{\sigma^{theor} = \sigma_0 \sqrt{1 + \alpha \left( 1 + w \right) e^{-w}}}
\label{equ314}
\end{equation}

For different combinations of $(\alpha, w)$,  we calculate $\sigma^{theor}$, and then use it for the FP fit.
As it can be seen from Eq. (\ref{equ314}), the Newtonian case can be recovered for  $\alpha = 0$ or $w = -1$, and then $\sigma^{theor} = \sigma_0$. However, $w$ has to be a positive number, and then $\alpha \approx 0$.

\subsection{The 3D fit of  fundamental plane  to  the observations}

Considering Eq.\eqref{equ11}, let us perform a 3D fit of the function $\log(r_e)$, depending on the two independent variables $\log(\sigma^{theor})$ and $\log(I_e)$ vs the observational data. We adopt a  least-squares algorithm. See \cite{bork16,bork19} for more details. In this way, we obtain the best fit coefficients of the FP described by  Eq.\eqref{equ11}, that is  $a$, $b$ and $c$.

\section{Constraining the Yukawa parameters}
\label{sec4}

\begin{figure}[ht!]
\centering
\includegraphics[width=0.48\textwidth]{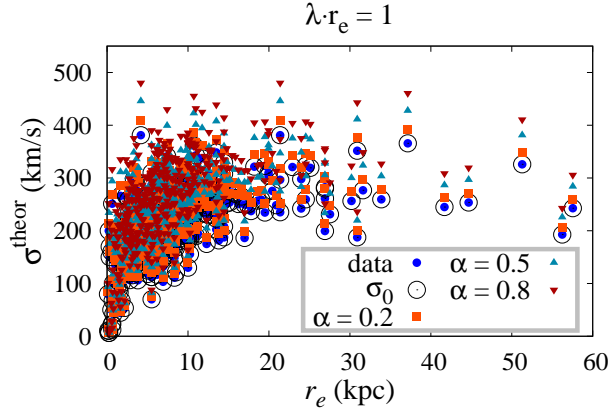}
\caption{Velocity dispersion $\sigma$ as a function of the effective radius $r_e$, for a sample of elliptical galaxies listed in Table 1 of \cite{burs97}. The observed values (blue full circles) and the Newtonian velocity dispersion at the effective radius $\sigma_0$ are taken from \cite{burs97}. Theoretical values of velocity dispersion $\sigma^{theor}$ are calculated for the following product of Yukawa parameter $\lambda$ and effective radius $r_e$: $\lambda \cdot r_e$ = 1 and for the three values of Yukawa  parameter $\alpha$: 0.2, 0.5 and 0.8.}
\label{fig01}
\end{figure}

\begin{figure}[ht!]
\centering
\includegraphics[width=0.48\textwidth]{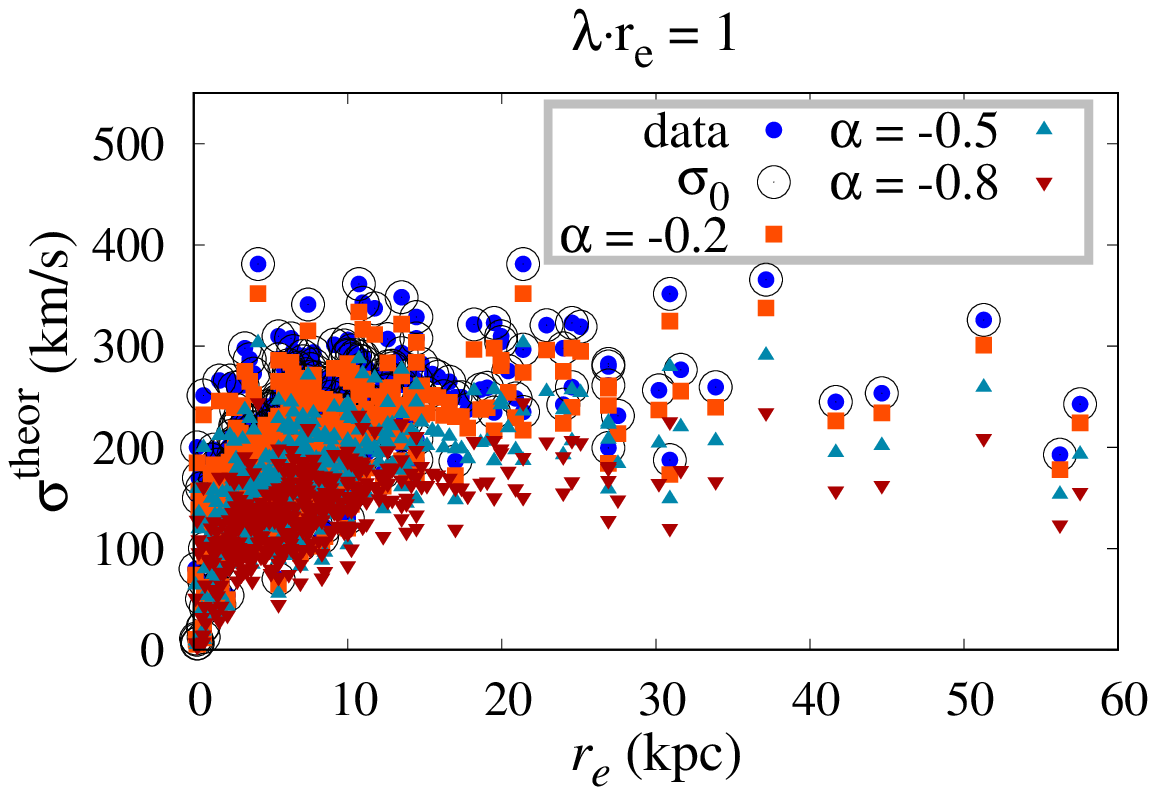}
\caption{The same as in Fig. \ref{fig01}, but for negative values of $\alpha: -0.2, -0.5, -0.8$.}
\label{fig02}
\end{figure}

\begin{figure*}[ht!]
\centering
\includegraphics[width=0.85\textwidth]{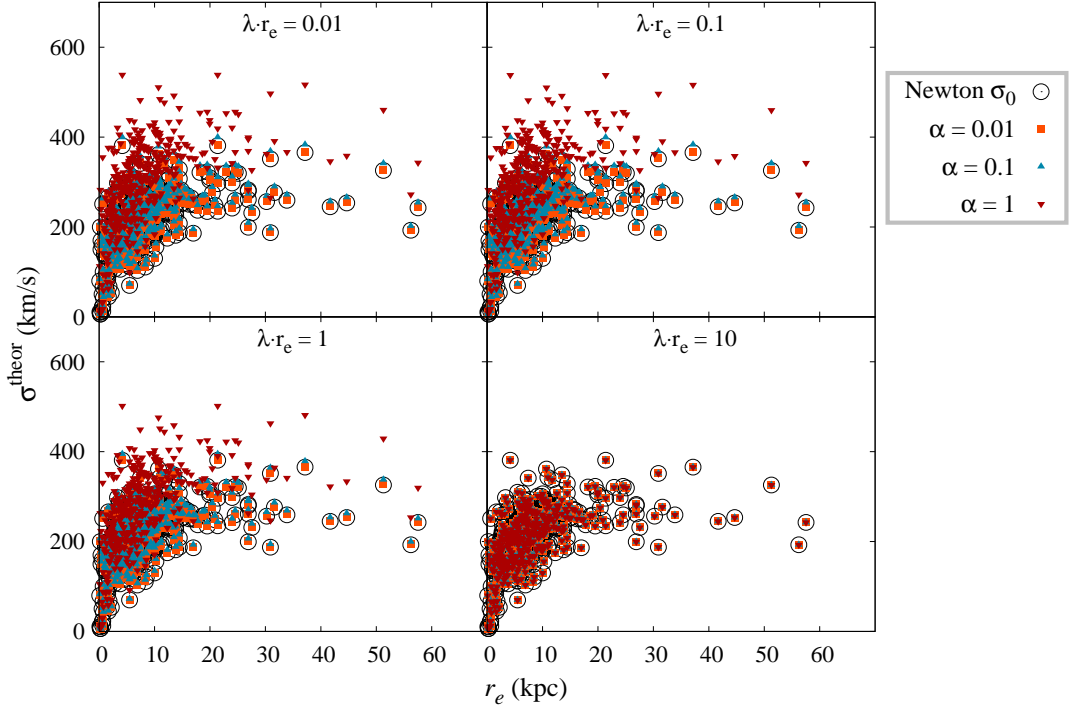}
\caption{Velocity dispersion $\sigma$ as a function of the effective radius $r_e$ for elliptical galaxies, for four different values of the $\lambda \cdot r_e$ product: 0.01, 0.1, 1 and 10. The Newtonian velocity dispersion inside the effective radius, $\sigma_0$, is taken from \cite{burs97}, and the calculated velocity dispersion $\sigma^{theor}$ is obtained for the three values of the parameter $\alpha$: 0.01, 0.1 and 1.}
\label{fig03}
\end{figure*}

\begin{figure*}[ht!]
\centering
\includegraphics[width=0.85\textwidth]{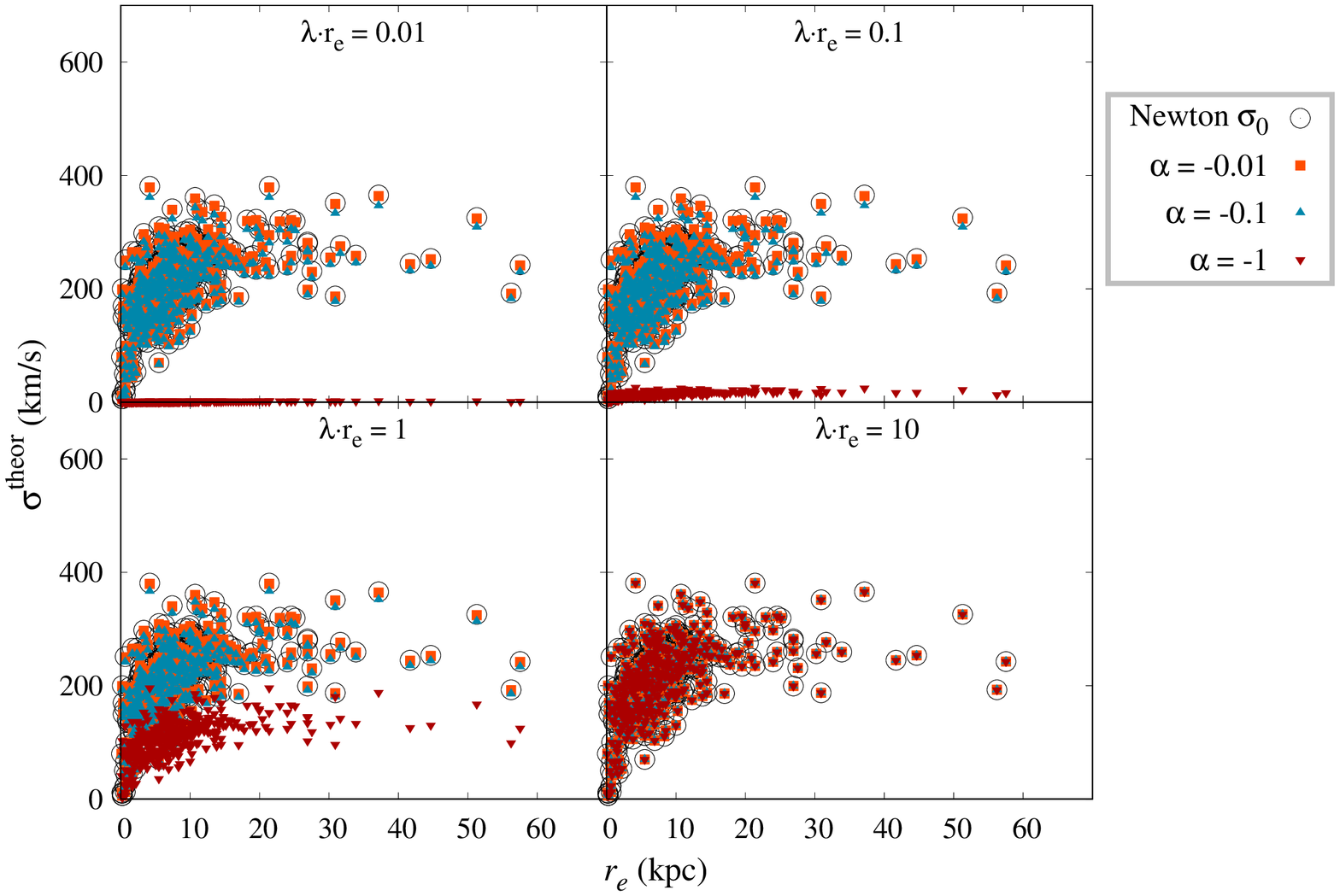}
\caption{The same as in Fig. \ref{fig03}, but for negative values of $\alpha: -0.01, -0.1, -1$.}
\label{fig04}
\end{figure*}

\begin{figure*}[ht!]
\centering
\includegraphics[width=0.85\textwidth]{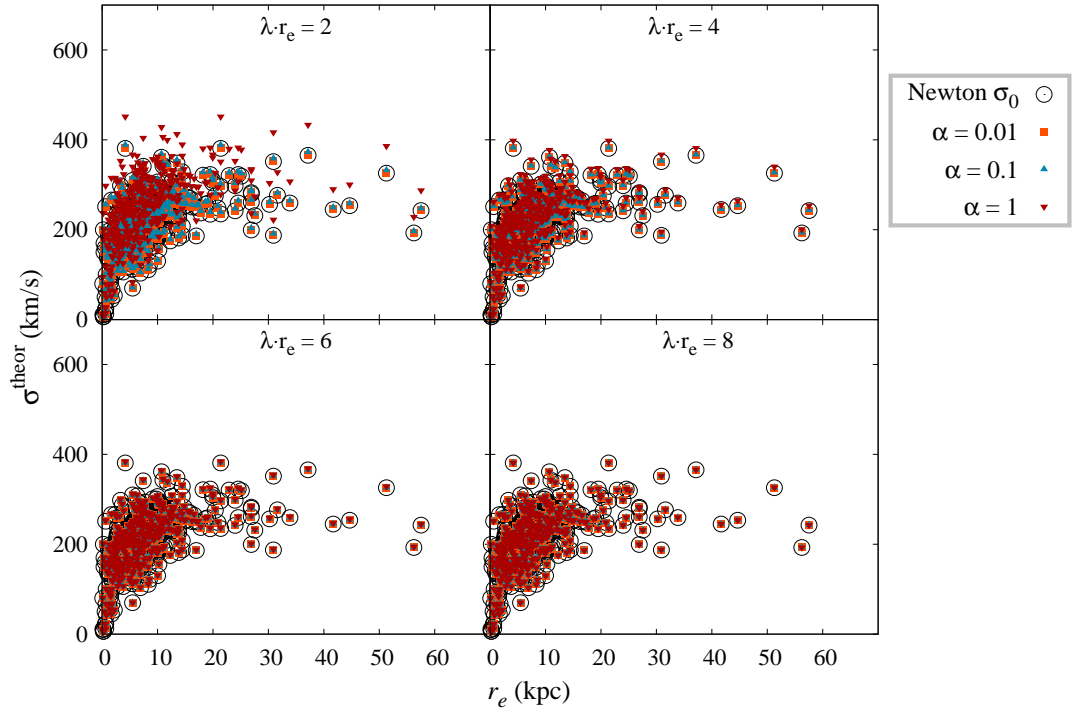}
\caption{The same as in Fig. \ref{fig03}, but for the following four $\lambda \cdot r_e$ products: 2, 4, 6 and 8.}
\label{fig05}
\end{figure*}

\begin{figure*}[ht!]
\centering
\includegraphics[width=0.85\textwidth]{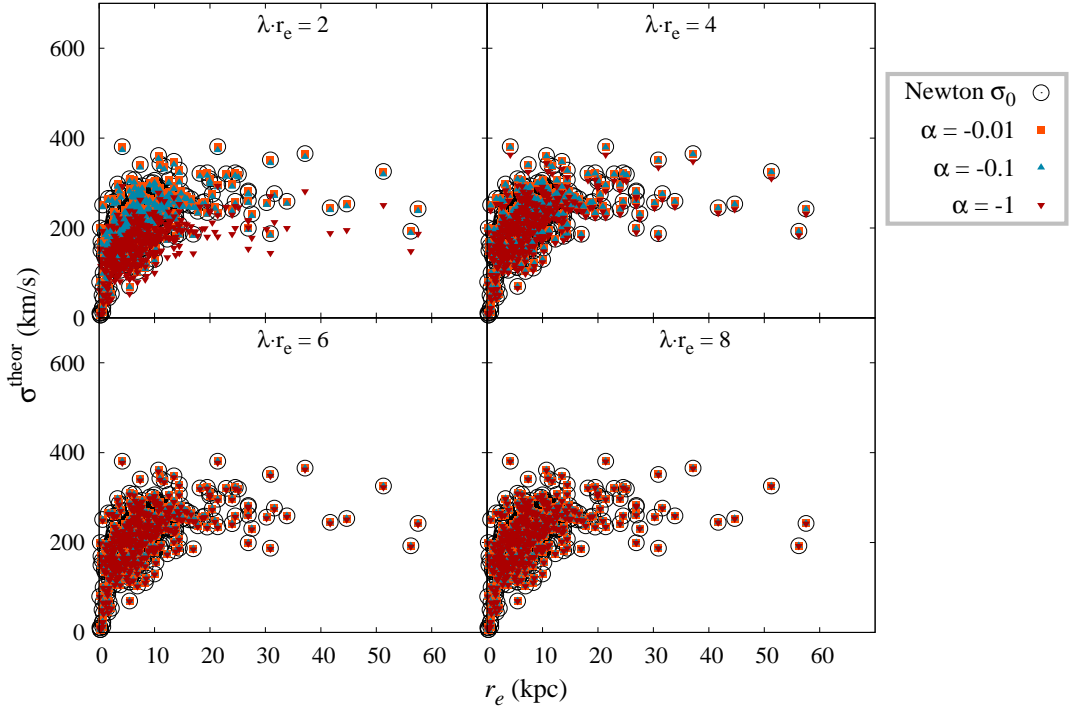}
\caption{The same as in Fig. \ref{fig05}, but for the following negative values of $\alpha: -0.01, -0.1, -1$.}
\label{fig06}
\end{figure*}

\begin{figure}[ht!]
\centering
\includegraphics[width=0.48\textwidth]{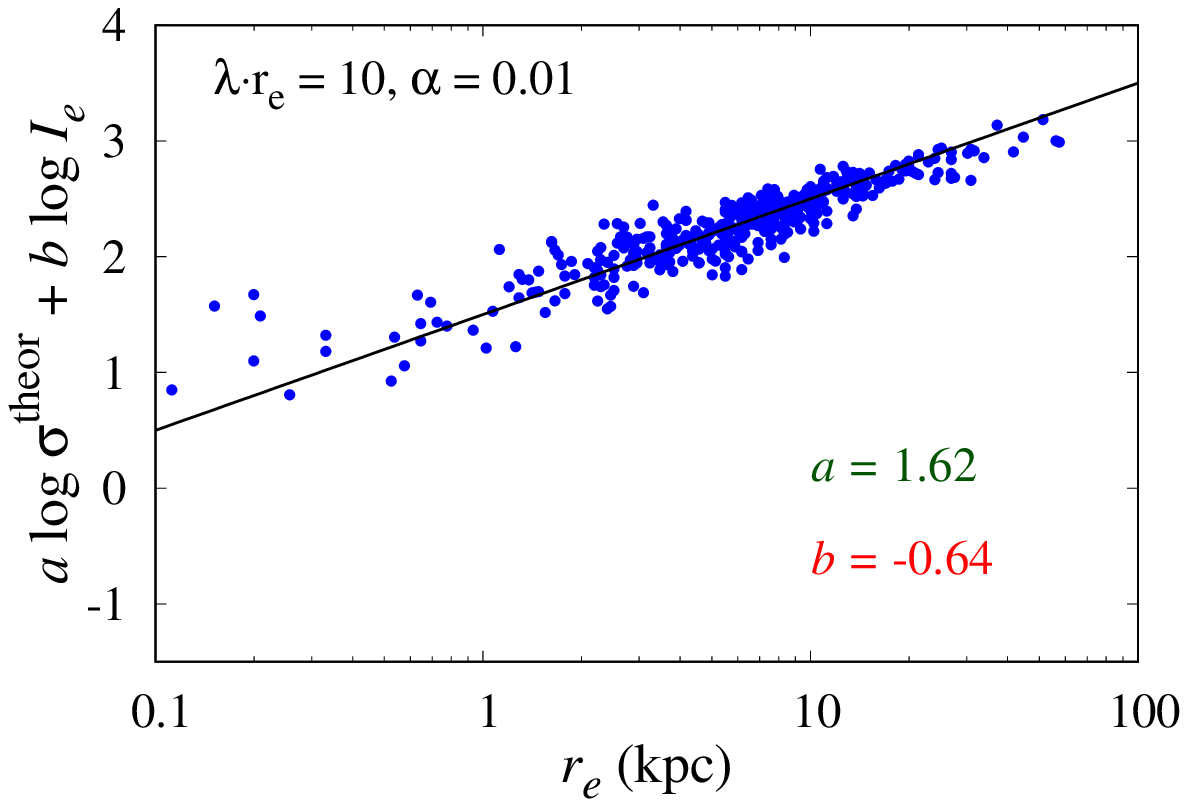}
\caption{Fundamental plane of elliptical galaxies with calculated velocity dispersion $\sigma^{theor}$,  observed effective radius $r_e$ and observed mean surface brightness (within the effective radius) $I_e$. For a given pair of Yukawa  parameters $(\lambda, \alpha$), we derive the  FP coefficients ($a$, $b$): $a = 1.62$, $b = -0.64$. Black solid line represents the best 3D fit of FP.}
\label{fig07}
\end{figure}

Let us now  vary the parameters $\lambda$ and $\alpha$ and discuss the matching between theoretical results and observations for velocity dispersion $\sigma$ as a function of the effective radius $r_e$ in the  case of elliptical galaxies listed in Table 1 of Ref. \cite{burs97}. First, we  fix  the value $\lambda \cdot r_e$ = 1 and vary parameter $\alpha$. Then, we  try different combination of values $\lambda \cdot r_e$ and $\alpha$. Finally, we will  discuss the results.
 
In Fig.\ref{fig01}, we show velocity dispersion $\sigma$ as a function of effective radius $r_e$. Theoretical values of velocity dispersion $\sigma^{theor}$ are presented for $\lambda \cdot r_e$ = 1 and for the three values of Yukawa  parameter $\alpha$: 0.2, 0.5 and 0.8. From our results,  we can see that the agreement between observed and theoretical values is relatively good only in case of small values of parameter $\alpha$ ($\alpha$ = 0.2). For larger values of parameter $\alpha$ ($\alpha$ = 0.5 or 0.8) agreement is very poor. 

In Fig.\ref{fig02}, it is shown the  velocity dispersion $\sigma$ as a function of the effective radius $r_e$, but for negative values of $\alpha: -0.2, -0.5, -0.8$ and for $\lambda \cdot r_e$ = 1. Like in the previous case, we can conclude that the agreement between observed and theoretical values is relatively good only in case of small values of parameter $\alpha$, that is  $\alpha$ = -0.2. For larger values of parameter $\alpha$ ($\alpha$ = -0.5 or -0.8) agreement is very poor. 

In Fig.\ref{fig03},  velocity dispersion $\sigma$ is shown as a function of effective radius $r_e$, but for four different values of the $\lambda \cdot r_e$ products: 0.01, 0.1, 1 and 10 and for  three different values of parameter $\alpha$: 0.01, 0.1 and 1. We can see that, for  $\lambda \cdot r_e$ = 10,  the agreement between observed and theoretical values is excellent for all  the three studied values of $\alpha$. If we look at Eq. (\ref{equ314}), $w$ = $\lambda \cdot r_e$, $e^{-w}$ is a very small number and, even if $\alpha$ is close to 1, $\sigma^{theor}$ is  close to $\sigma_0$. This is the reason why when $\lambda \cdot r_e$ = 10, the agreement between observed and theoretical values is excellent for all the three studied values of $\alpha$. In the case when $\alpha$ = 1, agreement for $\lambda \cdot r_e$ =  0.01, 0.1, 1 is poor, but it is better in the case of smaller values of $\lambda \cdot r_e$ ($\lambda \cdot r_e$ = 0.01, 0.1). In cases of small values of parameter $\alpha$ = 0.01 and 0.1,  agreement is very good for all the three different values of $\lambda \cdot r_e$. ($\lambda \cdot r_e$ =  0.01, 0.1, 1). 

In Fig.\ref{fig04},  we show velocity dispersion $\sigma$ as a function of effective radius $r_e$, for the same values of the $\lambda \cdot r_e$ product: 0.01, 0.1, 1 and 10, like in 
Fig.\ref{fig03}, but for the three different negative values of parameter $\alpha$: -0.01, -0.1 and -1. We  notice that $\alpha$ can take positive and negative values and the dependence of results is not symmetric with respect to the  sign of $\alpha$. From Figs.\ref{fig03} and \ref{fig04}, we can conclude that, in both cases, if value of $\alpha$ is small enough  ($\lesssim 0.1$) agreement between observed and theoretical values is excellent.

Figs.\ref{fig05} and \ref{fig06} show the same like Figs.\ref{fig03} and \ref{fig04}, but for the following four values of the $\lambda \cdot r_e$ product: 2, 4, 6 and 8 (now, scale is much more narrow then in Figs.\ref{fig03} and \ref{fig04}).  Figs.\ref{fig05} and \ref{fig06} help us to find when the $\lambda \cdot r_e$ product is enough large to give satisfactory agreement between observed and theoretical values. We can conclude that it happens when the value of the product $\lambda \cdot r_e$ is $\gtrsim 6$.

In Fig.\ref{fig07},  we present the FP of ellipticals with velocity dispersion $\sigma^{theor}$, observed effective radius $r_e$ and observed mean surface brightness $I_e$. For a given pair of Yukawa  parameters $(\lambda, \alpha$) (we choose $\lambda \cdot r_e$ =  10 and $\alpha$ = 0.01), using the same procedure described in \cite{bork16},  we calculate FP coefficients ($a$, $b$), and  obtain the values $a = 1.62$, $b = -0.64$. We perform now the fitting procedure (see \cite{bork16} for details) and obtain the black solid line which represents the best 3D fit of FP. Our calculated FP coefficients ($a$, $b$) are in good agreement with FP coefficients obtained by Bender et al.  using observational data \cite{bend92}. 
In other words, the thickness and the tilt of FP, derived from observations, fix the  values of  $\alpha$ and $\lambda$ in Eq.\eqref{equ21}. For the specific case of $f(R)$ gravity, we can infer the values of $f'_0$ and $f''_0$, that is, we can  reconstruct, up to the second order, the polynomial in  Eq.\eqref{sertay}. Analogue procedures can be developed for any model in Table 1.

\section{Discussion and conclusions}
\label{sec5}

Considering  results  in  Figs.\ref{fig01}-\ref{fig07},  we  conclude that the Yukawa-like correction can have an important influence on stellar dynamics of ellipticals, and hence on their FP.  These results also show that some of the theories of gravity reported in Table 1 are applicable only at some specific domains for the scale 
distance $L$. For example, the $f(R,R_{\alpha,\beta} R^{\alpha,\beta)}$ 
theory (4th case in Table 1) should be more suitable for 
shorter scales (i.e. for smaller values of $L$), since the coefficients 
in its potential ($\frac{1}{3}$ and $-\frac{4}{3}$) are not sufficiently 
small in magnitude, as it is expected for such scales.  A similar conclusion holds for the modified Gauss-Bonnet 
gravity \cite{nojiri} (see the  5th case from Table 1). This result  indicates that 
characteristic scale lengths of these two theories are possibly related 
to the effective radii $r_e$ of the ellipticals, as it is for 
$f(R)=R^n$ gravity (see \cite{bork16}). Due to opposite signs of the two 
exponential terms in the potential, this is valid only if their scale 
lengths are approximately the same. However, if this is not the case, by a
suitable choice of  scale lengths, it is possible to obtain a good 
agreement with observations at different astronomical and cosmological 
scales.
In this sense, FP can be a useful tool to select reliable theories.

Specifically, we can compare the Yukawa correction of the present  investigation with corrections in our previous paper \cite{bork16}, where we recovered  the FP using a power-law $f(R)=R^n$ gravity. There, the best agreement with the astronomical observations is obtained for values $r_c/r_e$ = 0.001 - 0.01. That is why we can say that our further gravitational radius $r_c$ is more important than $r_e$ to model stellar dynamics. In that case, the gravity parameter $\beta$ (values for $\beta$ that we analyzed: 0.02; 0.04; 0.06 and 0.08) gave relatively good results. There 
$\beta$ is a parameter related to the correction of Newtonian potential; it is  function of the index $n$ of a given $f(R)$ gravity model. We obtained a strong dependence of results on values $r_c/r_e$ and a weak dependence on the  values of $\beta$. In the case of the Yukawa correction,  the best agreement is obtained for $w$ = $\lambda \cdot r_e$ $\gtrsim 6$ (for all the three studied values of $\alpha$). This is the same tendency like in our previous study when we obtained the best agreement for values $r_c/r_e$ = 0.001 - 0.01. However, in this case, we can get also very good agreement even for much smaller values of $w$ = $\lambda \cdot r_e$, but only if $\alpha$ is small enough in  magnitude (i.e. $\lesssim 0.1$).  We can conclude that $\alpha$ can take positive and negative values and the dependence of results is not symmetric with respect of sign of $\alpha$, but in both cases, if value of $\alpha$ is small enough in magnitude, agreement is very good. The interpretation of this result is that we have to expect that Yukawa-like correction is relatively small with respect to the Newtonian leading term but, at long distances with respect to $r_e$, corrections can become relevant. In other words, for scale distances  $L=\lambda^{-1}\simeq 10 r_e$ Yukawa corrections become important and fit the distribution of ellipticals on the FP.

As a final consideration, it is worth saying that $\alpha$ and $\lambda$ are related to the given class of theories. As discussed above, they are related to $f'_0$  and $f''_0$ for $f(R)$ gravity. This means that the shape of $f(R)$ function is constrained by the FP parameters. The same can be done for the other theories in Table 1. This will be the argument of a forthcoming paper. 

\paragraph{Acknowledgments}
We wish to acknowledge the support by the Istituto Nazionale di Fisica Nucleare, Sezione di Napoli, Italy, iniziative specifiche QGSKY and MOONLIGHT-2 (S.C.), and Ministry of Education, Science and Technological Development of the Republic of Serbia (V.B.J., D.B. and P.J.). The authors also acknowledge the support of the COST Action CA15117 (CANTATA), supported by COST (European Cooperation in Science and Technology).

\end{document}